\documentclass[aps,prl,twocolumn,showpacs,preprintnumbers]{revtex4-1}

\usepackage{psfrag,graphicx}
\usepackage{dcolumn}
\usepackage{amsmath,amssymb}
\usepackage{bm}
\usepackage{amsfonts,amssymb,amsmath}        % for math symbols.
\usepackage{epstopdf}

\newcommand{\be}{\begin{equation}}
\newcommand{\ee}{\end{equation}}
\newcommand{\bq}{\begin{eqnarray}}
\newcommand{\eq}{\end{eqnarray}}

\newcommand{\ket}[1]{\left | \, #1 \right\rangle}

\begin{document}

\title{High Threshold Error Correction for the Surface Code}
\author{James R. Wootton and Daniel Loss}
\affiliation{Department of Physics, University of Basel, Klingelbergstrasse 82, CH-4056 Basel, Switzerland}

\begin{abstract}

An algorithm is presented for error correction in the surface code quantum memory. This is shown to correct depolarizing noise up to a threshold error rate of $18.5 \%$, exceeding previous results and coming close to the upper bound of $18.9 \%$. The time complexity of the algorithm is found to be polynomial with error suppression, allowing efficient error correction for codes of realistic sizes.

\end{abstract}


\maketitle

\emph{Introduction:} Topological error correcting codes, and the topological quantum computation that they may be used for, have attracted wide attention in recent years \cite{double,dennis,freedman,raussendorf,pachos}. As such, it is important to determine the threshold error rates for realistic error models and find efficient error correction algorithms to achieve them. The most studied, and most realistic topological error correcting codes are the surface codes \cite{double,dennis}, and the most realistic error model that is well-studied is that of depolarizing noise. The application of this noise model to a surface code induces correlations between different kinds of topological defects. Thus far, error correction algorithms have only been found that correct up to an error threshold of $16.5 \%$, the upper bound achievable when the correlations are ignored \cite{blossom,fowler,poulin}. Here we present an efficient algorithm that can correct beyond this bound. A threshold of $18.5 \%$ is found, falling only a little short of the recently established $18.9 \%$ limit \cite{bombin,ozheki}.

\begin{figure}[t]
\begin{center}
{\includegraphics[width=6cm]{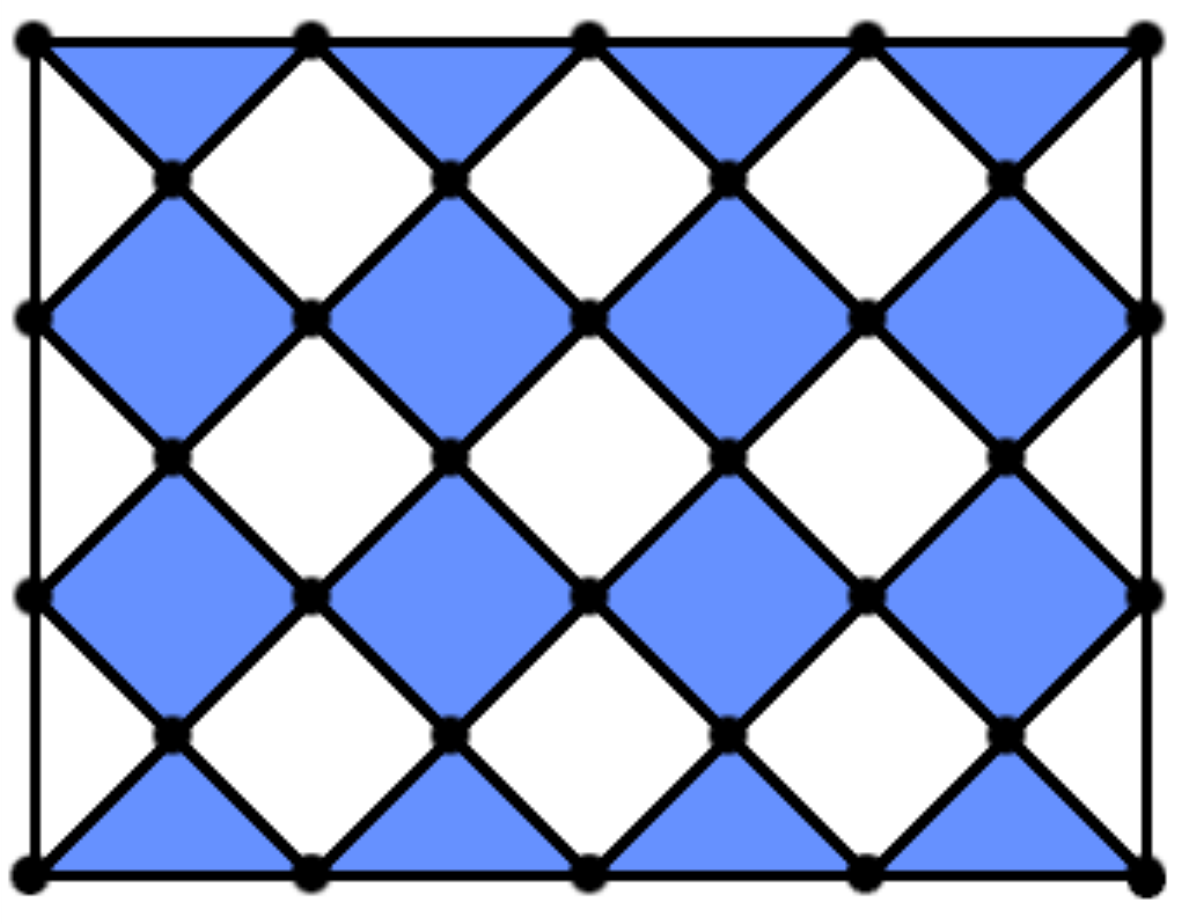}}
\caption{\label{fig1} The spin lattice on which an $L \times L$ planar code is defined, with $s$-plaquettes shown in blue and $p$-plaquettes in white. A spin-$1/2$ particle resides on each vertex. The linear size $L$ is characterized by the number of $s$-plaquettes along each side, with $L=4$ in this case.}
\end{center}
\end{figure}

\emph{The planar code:} The algorithm presented below is designed to correct errors in the the planar code, the planar variant of Kitaev's surface codes \cite{double,dennis}. The code is defined on the spin lattice of Fig. \ref{fig1}, where a spin-$1/2$ particle is placed on each vertex. The following Hermitian operators are then defined around each plaquette of the lattice,
\be
A_s = \prod_{i \in s} \sigma^x_i, \,\,\, B_p = \prod_{i \in p} \sigma^z_i.
\ee
These operators determine the anyonic occupation of their corresponding plaquettes, with so-called flux anyons on the $p$-plaquettes and charge anyons on the $s$-plaquettes. Since the operators mutually commute, they also form the stabilizers of a stabilizer code. The anyonic vacuum is the corresponding stabilizer space and the anyon configuration is the syndrome. The code can store a single qubit, whose state is determined by the anyonic occupations of the edges. The $X$ ($Z$) basis of the stored qubit may be chosen such that the $\ket{+}$ ($\ket{0}$) state corresponds to the vacuum on the top (left) edge and $\ket{-}$ ($\ket{1}$) corresponds to a flux (charge) anyon. The effect of errors on the spins is to create and move anyons, causing logical errors when they are moved off the edges.

\emph{Depolarizing noise:} The error model considered in this study is that of single qubit depolarizing noise. This is characterized by an error rate, $p$, which is taken to be equal for all spins. The probability that no error occurs on a spin is $1-p$. Otherwise, a $\sigma^x$, $\sigma^y$ or $\sigma^z$ error is applied, each with probability $p/3$. Such noise therefore takes an arbitrary single qubit  state $\rho$ and transforms it to,
\be
D_p (\rho) = (1-p) \rho + \sum_{\alpha=x,y,z} (p/3) \sigma^\alpha \, \rho \, \sigma^\alpha.
\ee
In the planar code, such noise results in correlations between the configurations of charge and flux anyons. Should these be ignored, error correction can be achieved so long as the probability that either a $\sigma^x$ or a $\sigma^y$ error occurs (or equivalently a $\sigma^z$ or a $\sigma^y$ error) is less than around $11\%$ \cite{dennis}. This gives a threshold of $p_c \approx 16.5 \%$. If the correlations are taken into account, the threshold increases to $p_c \approx 18.9 \%$ \cite{bombin,ozheki}.

\emph{Error correction:} Suppose a planar code, initially prepared in a state of the stabilizer space, is subject to depolarizing noise with a known rate $p$. Given the resulting anyon configuration (measurement of which we assume to be perfect), the job of error correction is to determine which of the four possible logical errors was caused by the physical errors.

Let us use $e$ to denote a configuration of errors, which records whether $\openone$, $\sigma_x$, $\sigma_y$ or $\sigma_z$ has occurred on each physical spin. Let us also use $A$ to denote a configuration of anyons and $E$ to denote the logical error ($\openone$, $X$, $Y$ or $Z$) that has occurred on the encoded qubit. Each $e$ is consistent with a unique $A$ and a unique $E$, so let us also use $A$ and $E$ to denote the set of $e$ consistent with the anyon configuration $A$ and logical error $E$, respectively. Given an anyon configuration $A$ after measurement of the stabilizers, the probability for each logical error is,
\be
P(E|A) = \sum_{e \in A \cap E} P(e|A).
\ee
Here $P(e|A)$ is the probability that the error configuration $e$ occurred given that the anyon configuration is $A$, etc. It can then be assumed that whichever $E$ is most likely is that which occurred, and error correction can be performed accordingly. For any $p<p_c$, this error correction procedure succeeds with a probability that tends to unity as  $L\rightarrow\infty$. For $p>p_c$ the success probability tends to $1/2$ in this limit, making error correction no better than guessing.

Note that $P(e|A)$ can be related to the unconditioned probabilities of $e$ and $A$ by $P(e|A)=P(e)/P(A)$. Since $P(A)$ is a common factor for all $E$, it does not need to be calculated in order to determine which of the $P(E|A)$ is greater, and hence which $E$ is most likely. The $P(e)$ may be calculated easily. For depolarizing noise $P(e) = (1-p)^{1-n_e} (p/3)^{n_e}$, where $n_e$ is the number of spins on which a $\sigma_x$, $\sigma_y$ or $\sigma_z$ has occurred on the error configuration $e$. The number of error configurations consistent with any anyon configuration is $2^N$, where $N=2L^2-1$ is the total number of plaquettes in the code. Calculating the $P(E|A)$ using a brute force approach will therefore take a time that is exponential with the system size. In fact, the scaling of this is so bad that no existent computer could correct an $L=11$ planar code in less than the age of the universe. As such, approximate methods are used to determine the most likely logical error for any anyon configuration. These achieve thresholds that are lower than the ideal case, but run for realistic time-scales \cite{blossom,fowler,poulin}.

The algorithm presented here uses a Markov chain Monte Carlo method to sample error configurations from the distribution $P(e|A)$. By taking many such samples, the probabilities $P(E|A)$ may then be approximated and hence the most likely logical error found. The most straightforward way to carry out this procedure, given an anyon configuration $A$, is using the Metropolis method as follows \cite{metropolis}. First a pattern of errors $e_0 \in A$ is generated randomly. This can be done in $O(L^2)$ time by first placing errors such that all anyons are connected, and then randomly applying each of the stabilizer. The first step ensures that $e_0$ is within $A$. The second ensures that it is random, since application of stabilizers deforms the error configuration without changing the anyon configuration. Once $e_0$ is generated, it can be used to generate a second configuration, $e_1 \in A$. To do this, a random change is made to $e_0$ to create a configuration $e_0' \in A$. The ratio,
\be \label{r}
r(e_0,e'_0) = \frac{P(e_0'|A)}{P(e_0|A)} = \left( \frac{p/3}{1-p} \right)^{n_{e_0'}-n_{e_0}} ,
\ee
is then determined. If $r(e,e')>1$, we set $e_1 = e_0'$. Otherwise we set $e_1 = e_0'$ with probability $r(e,e')$ and $e_1 = e_0$ with probability $1-r(e,e')$. This process then continues until the sequence of $e_n$ converges, at which point they will be generated according to the distribution $P(e|A)$ \cite{metropolis}.
% typo corrected

The most intuitive method that could be used to generate each $e_n'$ from each $e_n$ is to randomly pick a stabilizer and apply it. This will cause an $O(1)$ change in the number of errors and hence yield an $r(e_n,e'_n)$ of $O(1)$. However, only making such changes means that only error configurations corresponding to the same $E$ as $e_0$ will be generated. Additional changes in which logical operators spanning the code can be randomly applied must therefore also be made, such that configurations from all $E$ are sampled from. However, these will add $O(L)$ errors to any configuration on which they are applied, resulting in $r(e_n,e'_n) = O(\exp -L)$. Since the acceptance of such changes is exponentially small, the time taken to convergence will be at least $O(\exp L)$. Some additional methods are therefore required to avoid this source of inefficiency.

A solution to the problem is to use parallel tempering \cite{temp}. For this, many Markov chains such as that described above are run in parallel. Let us use $N_c$ to denote the number of such chains, and restrict it to being odd. The first chain (which we will refer to as the bottom chain) works exactly as described above. Each $e'_n$ is generated from $e_n$ by application of a random stabilizer. No logical operators are applied to change the value of $E$. The second chain works in the same way, except for a difference in the calculation of the $r(e_n,e'_n)$. Instead of using the error rate $p$ when calculating the $P(e)$, a slightly higher error rate $p_2 = p+\Delta$ is used, where $\Delta = (0.75-p)/(N_c-1)$. Similarly the $m$th chain will use an error rate of $p_m = p+(m-1) \Delta$. Using this prescription, the $N_c$th chain (which we will refer to as the top chain) has $p_{N_c} = 0.75$, and so $r(e_n,e'_n)=1$ in all cases. As such, we need not restrict each $e_n'$ for this chain to be only an $O(1)$ change away from $e_n$. Accordingly, the $e_n'$ are generated randomly and independently from the $e_n$ by randomly applying all stabilizers and logical operators. It is therefore in the top chain, and only the top chain, where the value of $E$ changes.

The randomness in $E$ generated in the top chain is introduced to the rest of them as follows. After running each chain for a certain number of iterations, swaps between neighbouring chains are attempted. For a swap between chains $m$ and $m+1$, the ratio 
\be
r(e^m,e^{m+1}) = \left( \frac{p_m}{p_{m+1}}\frac{1-p_{m+1}}{1-p_{m}} \right)^{n_{e^{m+1}}-n_{e^{m}}},
\ee
is calculated. Here $e^m$ denotes the current state of the $m$th chain, etc. This is a straightforward generalization of Eq. \ref{r} to the state of two chains rather than one, where the proposed change is the swap of states. If $r(e^m,e^{m+1})>1$, the configuration $e^m$ is set as the new state of the $m+1$th chain, and vice-versa. Otherwise this is done with probability $r(e^m,e^{m+1})$ and the chains are left alone with probability $1-r(e^m,e^{m+1})$. The Metropolis process is then again run on each chain for a number of iterations before a further break in which swaps are attempted, continuing until convergence. Henceforth we will refer to a certain number of Metropolis iterations followed by a break to attempt swaps as a `step' of the algorithm.

In order for the states of high chains to be able to migrate down to the bottom in a time faster than $O(\exp L)$, it must be ensured that the $r(e^m,e^{m+1})$ do not decay with system size. Since a system of side length $L$ has $2L^2$ physical spins, and since the number of errors in any chain should be proportional to its error probability, we see that the difference in the number of errors for two neighbouring chains is $n_{e^{m+1}}-n_{e^{m}} = O(L^2 \Delta)$. Also, if $\Delta$ is small, $\ln ([p_m/p_{m+1}][(1-p_{m+1})/(1-p_m)]) = O(\Delta)$. This means $r(e^m,e^{m+1}) = O(\exp[L^2 \Delta^2])$, and so $\Delta = O(L^{-1})$ will lead to $r(e^m,e^{m+1}) = O(1)$. In order to achieve this $N_c = O(L)$ chains are used. The numerical simulations confirm that this leads to $r(e^m,e^{m+1})$ that do not decay with system size.

The total number of unique samples originating in the top chain that have filtered down to the bottom is counted throughout the process as a measure of its progress. This number is denoted \texttt{tops0}. Convergence is tested for by a variant of the Geweke diagnostic \cite{geweke}. To do this the number of errors present in the first chain are recorded at the end of each step. Averages are then made over the second and fourth quarters of this data and these are compared. If the process has converged, these averages should be equal. As such, if the averages remain within a tolerance of $\epsilon$ of each other for a certain number of steps, the process is taken to be converged. This number of steps is taken to be that required for \texttt{tops0} to increase by an amount \texttt{SEQ}. To reduce serial correlations, and ensure that states from all chains have had a chance to migrate to the bottom, the comparison between the averages is not made until \texttt{tops0} has reached a value of \texttt{TOPS}. The values of $E$ are recorded during the period over which the averages remain within $\epsilon$. The logical error in the majority over all these is then taken to be the most likely.

The above tests for convergence of the process to its stationary distribution, $P(e|A)$. However, this is not necessarily required in order to determine which of the logical errors is most likely. As such, in addition to this first variant of the algorithm, we will consider also a second variant whose convergence test determines the point at which the most likely logical error becomes obvious. To do this, the value of $E$ is recorded at the end of each step and the majority values for the second and fourth quarters of this data are determined. If these remain equal for the number of steps required for \texttt{tops0} to increase by \texttt{SEQ}, their shared value is taken to be the most likely value of $E$. As before, to reduce serial correlations, the comparison between the averages is not made until $\texttt{tops0}=\texttt{TOPS}$. Also the values of $E$ are not recorded until $\texttt{tops0}=1$.

\emph{Results:} The task of an error correcting code is to reduce the logical error rate, $P$, to some desired value. The resources required for this task are the number of spins that must be used, and the time that is taken to decode the information at readout. In order for an error correction algorithm to be called efficient, it must be able to obtain any given $P$ with both a system size and run-time of $O(\rm{poly} \, P^{-1})$. Note that it is this scaling of the run-time of the algorithm with logical error rate that is of primary importance. The scaling of run-time with system size, which is often considered in studies of error correction algorithms, is only a secondary concern. In fact, since the planar code is theoretically capable of obtaining any given $P$ with system size $L = O(\log \, P^{-1} )$ (as long as the spin error rate is below threshold), the correction algorithm could have a run-time of $O(\exp \, L)$ and still be called efficient.

The algorithm was run according to the following procedure in order to determine its performance. First a pattern of errors was generated randomly according to the depolarizing noise model with a given error rate, $p$. The anyon configuration, $A$, corresponding to these errors was then measured and passed to the correction algorithm. The algorithm, which samples from  $p(e|A)$ according to the procedure of the previous section, then determines which logical error $E$ is most likely given the anyon configuration $A$. It is assumed that this is the logical error caused by the actual errors applied to the spins, and correction is performed accordingly. By comparing to the actual error configuration, the success of the correction is determined. By repeating the process for many samples, the success can be measured by calculating the probability that the error correction causes a logical bit-flip error, $P$.

\begin{figure}[t]
\begin{center}
{\includegraphics[width=1\columnwidth]{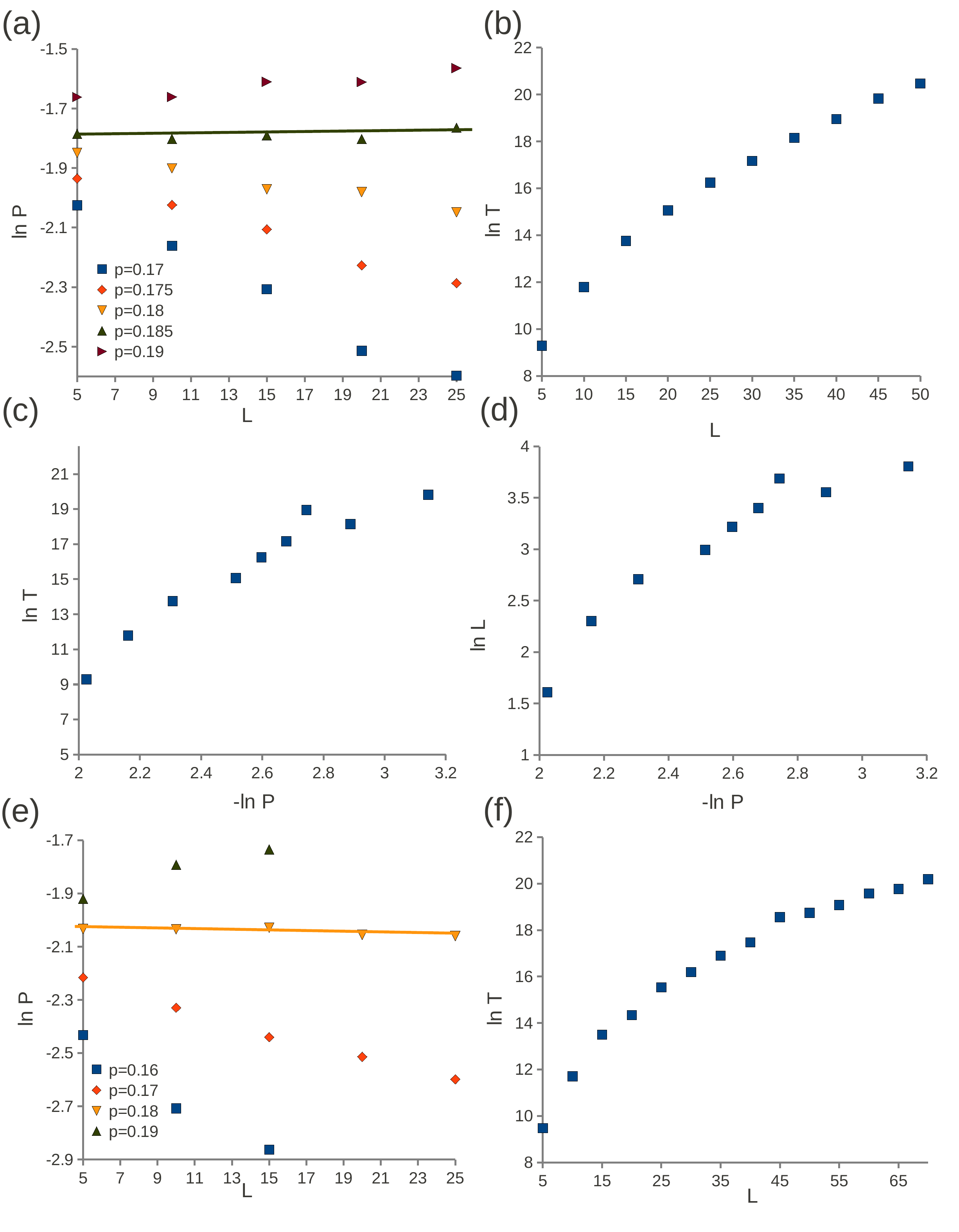}}
\caption{\label{fig2} Results for the first variant of the algorithm are presented in (a), (b), (c) and (d). Those for the second are in (e) and (f). Both were run using the nearest odd integer to $L$ for $N_c$, ten metropolis iterations per step, $\epsilon = 0.1$ and $\texttt{TOPS} = 10$. The first used $\texttt{SEQ} = 2$ and the second used $\texttt{SEQ} = 10$. (a) and (e) show plots of the logical bit flip error rate after correction, $P$, against linear system size, $L$. Each point was averaged over $10^4$ samples. Fit lines for the threshold values are shown as a guide to the eye. (b) and (f) show plots of $T$, the number of steps required by the algorithm before convergence, against $L$. Since only the order of magnitude of the run-time is important, the number of samples averaged over was reduced to between $10$ and $100$, allowing higher system sizes to be probed. The results in (c) and (d) are both for the first variant, and show what $T$ and $L$ are required to achieve a logical error rate $P$.}
\end{center}
\end{figure}

In Fig. \ref{fig2} (a) and (e) the logical bit flip error rate, $P$, is plotted against system size for a range of single spin error rates, $p$. If $p$ is under (over) the threshold value $p_c$ the logical error rate will decrease (increase) with system size. From the results, it is evident that $p_c \approx 0.185$ for the first variant and $p_c \approx 0.18$ for the second. These values fall slightly short of $p_c = 0.189$, the value that would be achieved by a brute force method \cite{bombin,ozheki}. Theoretically the algorithms should achieve the maximum value as $\epsilon \rightarrow 0$ or $\texttt{SEQ} \rightarrow \infty$. However, the runtime required for this will be prohibitive.

In Fig. \ref{fig2} (b) and (f) the time complexity of the two variants of the algorithm is considered as a function of system size. This uses the number of steps required by the algorithm before convergence, $T$, with data obtained for $p=0.17$. It is found that $\log T$ scales sublinearly with $L$, and hence $T$ scales subexponentially with $L$. Each step requires $O(\log L)$ actions on $N_c = O(L)$ chains. The former is required to generate the random numbers that choose which of the $O(L^2)$ stabilizers to apply during the Metropolis procedure. The the total time complexity of the process is therefore also subexponential in $L$. Note that the second variant has a more efficient scaling with system size than the first, but achieves lower logical error rates. This variant therefore balances an increase in efficiency with a decrease in effectiveness. This makes it more useful than the first variant in theoretical studies, since it allows higher system sizes to be probed, but less useful when actually performing error correction.

In Fig. \ref{fig2} (c) and (d) the time and system size complexity of the first variant are considered as a function of the acheived logical error rate, $P$. As well as a combination of the data in Fig. \ref{fig2} (a) and (b), logical error rates were also obtained for system sizes from $L=30$ to $L=45$ (also at $p=0.17$). Due to the prohibitive runtime for such sizes, this data was not averaged over a fixed number of samples, but instead run until a fixed number of logical errors occurred. The number used for this was $10$ (though the number of samples is much greater than this). The plots show that $\log T$ and $\log L$ grow no faster than linearly with  $-\ln P$, meaning $T$ and $L$ grow no faster than polynomially with $P^{-1}$. In fact, the logarithms seems to scale sublinearly with $-\ln P$, meaning a similarly sublinear scaling of $T$ and $L$ with $P^{-1}$.

It is also important to determine the effectiveness of the algorithm for low error rates, for which the logical error rates should scale as $O(p^{\lfloor (d+1)/2 \rfloor})$ for small $p$ \cite{fowlerunpub}. The distance, $d$ of the planar code is $L+1$ for bit flip errors and $L$ for phase flip errors. Numerical simulations for the performance of codes from $L=2$ to $L=4$ for error rates from $0.5\%$ to $3\%$ show a good fit to such scaling, implying that this algorithm does indeed allow the code to utilize its full distance.

\emph{Conclusions:} We have presented an algorithm for error correction in the planar code. It is demonstrated that this achieves thresholds higher than existing algorithms, approaching the theoretical bounds. The efficiency is shown to be polynomial with error suppression. This allows effective error correction for applications where the classical post-processing may be left until readout, such as a planar code quantum memory. The method on which this algorithm is based, Markov chain Monte Carlo, is not limited to the code and error model presented here. Our work therefore forms a foundation on which error correction algorithms for other topological codes and error models may be built. Future work will be dedicated to further development of the the algorithm, to increase both the threshold and the efficiency. This will allow it to be applied to cases for which classical post-processing cannot be postponed, such as quantum computation. Also, application of the algorithm to the case of noisy stabilizer measurements is sure to yield important results for this physically realistic case.

\emph{Acknowledgements:} The authors would like to thank Beat R\"othlisberger, Abbas Al-Shimary and Austin Fowler for valuable discussions and comments. This work was supported by the Swiss NF, NCCR Nano, NCCR QSIT, and DARPA.

\appendix

\section{Appendix: Comparsion with MWPM}

Though the threshold achieved by an algorithm is an important measure of its effectiveness, it is by no means the whole story. It is also important, below threshold, for an algorithm to reduce the logical error rate $P$ as much as possible. One way to measure this is to determine the minimum system size required for the logical error rate to become lower than the physical error rate. This is therefore the system size at which the error correcting properties of the code begin to take effect, allowing proof of principle experiments to demonstrate its power. This minimum size will, of course, depend on the physical error rate $p$. The effectiveness of an algorithm is then characterized by a curve of the minimum effective system size against $p$. The better an algorithm performs, the smaller the code can be made while still performing effective error correction, and so the lower the curve.

\begin{figure}[h]
\begin{center}
{\includegraphics[width=\columnwidth]{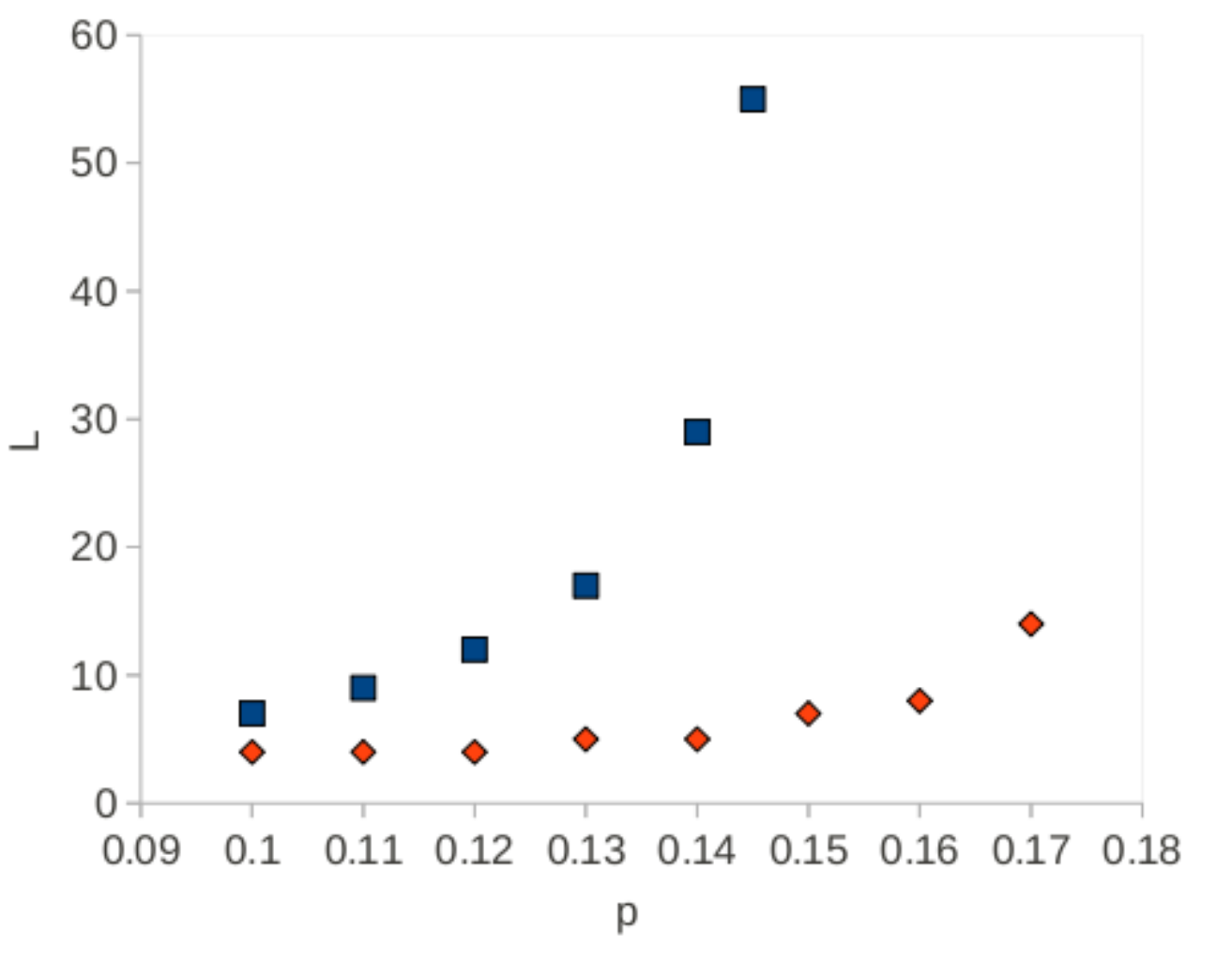}}
\caption{\label{fig3} Minimum system size, $L$, for which both logical bit and flip error rates are lower than their physical counterparts. Results for MWPM is shown are shown in blue, and those for the first variant of our algorithm are shown in red.}
\end{center}
\end{figure}

In Fig. \ref{fig3} this curve is shown for the first variant of the algorithm. For comparison, the curve for minimum weight perfect matching (MWPM) is also shown. The MWPM algorithm used does not employ a Delauney triangulation for increased efficiency, but instead performs the full matching. As such, it yields the most accurate result possible for the matching. Even so, our algorithm can be seen to perform much better. Its curve is consistently significantly lower, and so achieves effective error correction at much smaller system sizes.

Therefore, by doing nothing other than changing the method of classical post-processing, our algorithm will allow proof of principle experimental verification of topological error correction at higher error rates and with smaller system sizes. This provides a significant practical advantage.

\begin{thebibliography}{99}

\bibitem{double} A. Kitaev, Ann. Phys. (N.Y.) \textbf{303}, 2 (2003).
\bibitem{dennis} E. Dennis, A. Kitaev, A. Landahl, J. Preskill, J. Math. Phys. \textbf{43}, 4452 (2002).
\bibitem{freedman} M. H. Freedman, A. Kitaev, M. J. Larsen, and Z. Wang, Bull. Amer. Math. Soc., \textbf{40}, 31 (2003).
\bibitem{raussendorf} R. Raussendorf and J. Harrington, Phys. Rev. Lett. 98, 190504 (2007). 
\bibitem{pachos} G. Brennen and J. K. Pachos, Proc. R. Soc. London, A {\bf 464}, 2089 (2008).

\bibitem{blossom} J. Edmonds, Can. J. Math., \textbf{17}, 449 (1965).
\bibitem{fowler} D. S. Wang, A. G. Fowler, A. M. Stephens, L. C. L. Hollenberg, Quant. Inf. and Comp. \textbf{10}, 456 (2010).
\bibitem{poulin} G. Duclos-Cianci and D. Poulin, Phys. Rev. Lett. \textbf{104}, 050504 (2010).

\bibitem{bombin} H. Bombin, R. S. Andrist, M. Ohzeki, H. G. Katzgraber, M. A. Martin-Delgado, Phys. Rev. X \textbf{2}, 021004 (2012).
\bibitem{ozheki} M. Ohzeki, arXiv:1202.2593 Phys. Rev. A \textbf{85}, 060301 (2012).

\bibitem{metropolis} N. Metropolis, A. W. Rosenbluth, M. N. Rosenbluth, A. H. Teller, E. Teller, J. Chem. Phys., \textbf{21}, 1087 (1953).

\bibitem{temp} D. J. Earl and M. W. Deem, Phys. Chem. Chem. Phys., \textbf{7}, 3910 (2005).

\bibitem{geweke} M. K. Cowles and B. P. Carlin, J. Amer. Stat. Assoc., \textbf{91}, 883 (1996).

\bibitem{fowlerunpub} A. G. Fowler, A. C. Whiteside, L. C. L. Hollenberg, arXiv:1202.5602 (2012).
 
\end{thebibliography}
\end{document}